\documentclass[twocolumn,showpacs,preprintnumbers,amsmath,amssymb]{revtex4}

\usepackage{graphicx}
\usepackage{dcolumn}
\usepackage{bm}
\usepackage{color}
\usepackage{float}

\begin{document}


\title{Fusion of $^{7}$Li with $^{205}$Tl at near barrier energies}

\author{V. V. Parkar$^{1,2}$\footnote{vparkar@barc.gov.in}}
\author{Prasanna M.$^3$}
\author{Ruchi Rathod$^4$\footnote{Present address~:Physics Department, University of Notre Dame, Notre Dame, Indiana 46556, USA}}
\author{V. Jha$^{1,2}$}
\author{S. K. Pandit$^{1,2}$}
\author{A. Shrivastava$^{1,2}$}
\author{K. Mahata$^{1,2}$}
\author{K. Ramachandran$^{1}$}
\author{R. Palit$^6$}
\author{Md. S. R. Laskar$^6$}
\author{B. J. Roy$^{1,2}$}
\author{Bhushan Kanagalekar$^3$}
\author{B. G. Hegde$^3$}
\affiliation{$^1$Nuclear Physics Division, Bhabha Atomic Research Centre, Mumbai - 400085, India}
\affiliation{$^2$Homi Bhabha National Institute, Anushaktinagar, Mumbai - 400094, India}
\affiliation{$^3$Department of Physics, Rani Channamma University, Belagavi - 591156, India}
\affiliation{$^4$Sardar Vallabhbhai National Institute of Technology, Surat - 395007, India}
\affiliation{$^6$Department of Nuclear and Atomic Physics, Tata Institute of Fundamental Research, Mumbai - 400005, India}


\begin{abstract}
The complete and incomplete fusion cross sections for the $^{7}$Li+$^{205}$Tl reaction were measured at near barrier energies by online characteristic $\gamma$ ray detection technique. The complete fusion (CF) cross sections at energies above the Coulomb barrier were found to be suppressed by $\sim$ 26 \% compared to the coupled channel calculations. Reduced fusion cross sections for the present system at energies normalised to the Coulomb barrier were also found to be systematically lower than those with strongly bound projectiles forming a similar compound nucleus. The suppression observed in CF cross sections is found to be commensurate with the measured total incomplete fusion (ICF) cross sections. In the ICF cross sections, t capture is found to be dominant than $\alpha$ capture at all the measured energies. The systematic study of available CF, ICF and total fusion (TF) data with $^7$Li projectile is performed. 
\end{abstract}

\maketitle

\section{\label{sec:Intro} Introduction}
In the last several decades heavy ion fusion reactions have been studied to understand the interplay between nuclear structure and dynamics in colliding partners \cite{Becker88, Bala98, Dasgupta1998, Hagino12, Back2014, Mont17, Jha20, Jiang21}. Cluster transfer is also reported to play a key role in understanding the reaction mechanisms \cite{sanat21}. Heavy ion fusion reactions are important to extend the nuclear chart and for the synthesis of heavy elements. These reactions are also important for the understanding of the energy production and elemental synthesis in the stellar environments. With the advent of radioactive ion beams (RIBs), several novel exotic structures and phenomena like halo and Borromean nuclei have been discovered experimentally. Moreover, the experimental measurements performed with RIBs so far have shown that the fusion behaviour is uncertain and not so well understood.  It is expected  that the extended structure of these weakly bound nuclei may in principle lead a large enhancement of fusion. However, it can also be argued that due to their low binding energy, these nuclei may break up easily while approaching the fusion barrier, which may effectively reduce the complete fusion (CF) cross sections \cite{Keel09,Kola16}. Fusion with weakly bound stable projectiles such as $^{6,7}$Li and $^9$Be having cluster structures provide a good experimental ground to evaluate these contrasting propositions. 

It is now fairly well established  that in the  fusion reactions with weakly bound stable projectiles ($^{6,7}$Li and $^9$Be) on different targets the complete fusion, where the entire projectile or all its fragments are captured in the target, is suppressed when compared to predictions based on coupled channels calculations at energies above the Coulomb barrier \cite{Canto15, Jha20}.  Similarly, the reduction of CF cross sections has also been observed when the CF cross sections measured in reactions with weakly bound stable projectiles are compared with that measured with strongly bound projectiles forming similar compound nucleus. The suppression of  CF cross sections is often quantified through the CF suppression factor which has been \color{black} deduced \color{black} in several experiments with $^{6,7}$Li and $^9$Be projectiles on medium and heavy mass targets showing  a systematic behaviour \cite{Jha20}, which is found to be independent of target mass in many studies \cite{Gasq09,vvp10,Kundu16,Wang14}.  In general, the amount of suppression observed in CF cross sections is almost completely compensated by the incomplete fusion (ICF) cross sections, where part of the projectile is captured by the target. Although there is a large amount of CF data with weakly bound projectiles \cite{Jha20}, simultaneous measurement of CF and ICF data is available for only few systems. Recently, the damping of the nuclear shell effect with excitation energy has been studied by measuring neutron spectra following the triton transfer in $^{205}$Tl($^7$Li,$\alpha$)$^{208}$Pb reaction \cite{Rout13}. However the CF and ICF cross sections were not measured so far for this system. 

\color{black} In general, the suppression effects are expected to be dependent on the variation due to Coulomb and nuclear breakup and cluster transfer depending on the Z of the target. However,  the CF suppression with weakly bound projectiles $^{6,7}$Li and $^9$Be is found to be target independent.  In the present work, we have measured the CF and ICF cross sections for $^7$Li+$^{205}$Tl system and compared them with the similar existing data with $^7$Li projectile on several targets to understand the systematic behaviour. \color{black}

In this paper, we report the measurement of CF and ICF cross sections for $^{7}$Li+$^{205}$Tl system around the Coulomb barrier energies using online $\gamma$-ray detection technique. The available data of CF, ICF and TF cross sections with $^{7}$Li projectile on several targets was utilised to understand the systematic behaviour. The paper is organized as follows: the experimental details are described in Sec.~\ref{sec:Expt}. The experimental results, systematics of data along with statistical model and coupled channel calculations are given in Sec.~\ref{sec:Result}. The summary of the present study is given in Sec.~\ref{sec:Sum}.

\section{\label{sec:Expt} Experimental Details}
\begin{figure}
\includegraphics[width=110mm,trim=0.7cm 7.8cm 2.6cm 1.8cm, clip=true]{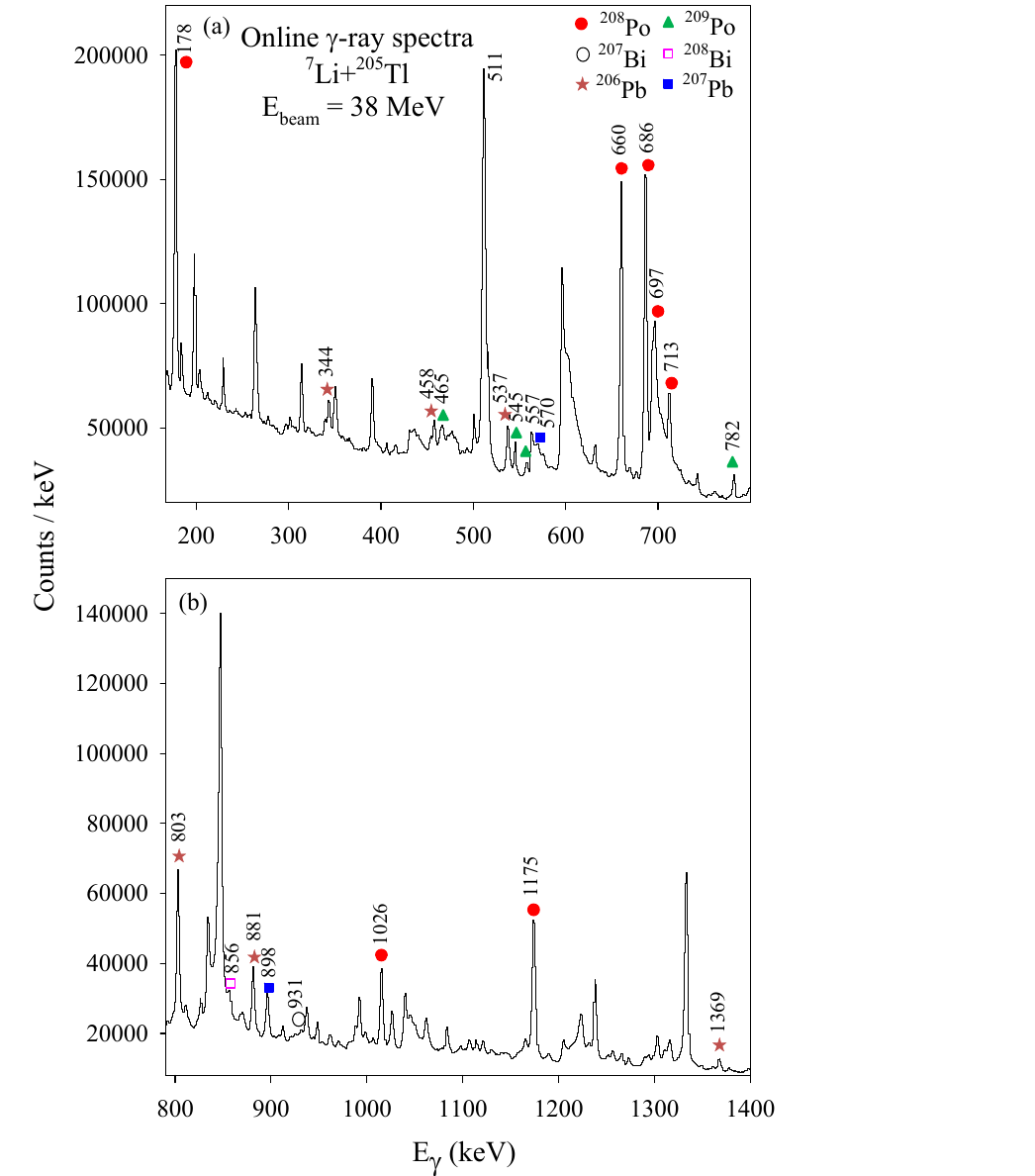}
\caption{\label{fig1} $\gamma$-ray add-back spectrum from all the clover detectors  obtained in $^7$Li+$^{205}$Tl system at E$_{\textrm{beam}}$ = 38 MeV. The $\gamma$ lines from the possible evaporation residues ($^{208,209}$Po) following CF are labeled. Also the $\gamma$ lines following the $\alpha$-capture channel ($^{207,208}$Bi) and t-capture channel ($^{206,207}$Pb) are marked.}
\end{figure}

The experiment was performed using $^{7}$Li beam from the BARC-TIFR Pelletron LINAC Facility, Mumbai, India at energies, E$_{\textrm{beam}}$ = 25, 26, 27, 28, 29, 31, 32.5, 34, 38 and 40 MeV. The target used was $^{205}$Tl of 1 mg/cm$^2$, evaporated on 25 $\mu$g/cm$^2$ carbon backing. Beam energies were corrected for the loss at half the target thickness and were further used in the analysis. Prompt $\gamma$-ray transitions were detected using nine Compton suppressed High Purity Germanium (HPGe) Clover detectors from Indian National Gamma Array (INGA) \cite{Pali12} surrounding the target chamber. In this array configuration, the detectors were arranged at three angles with three detectors each at $\pm$ 157$^{\circ}$, $\pm$ 140$^{\circ}$ and $\pm$ 90$^{\circ}$\color{black}. In addition, two Si surface barrier detectors with thicknesses 300 $\mu$m, acting as monitor detectors were placed at $\pm$ 25$^{\circ}$ for absolute normalisation purpose. The time stamped data were collected using a digital data acquisition system with a sampling rate of 100 MHz \cite{Pali12}. Efficiency and energy calibration of the clover detectors were carried out using standard calibrated $^{152}$Eu and $^{133}$Ba $\gamma$-ray sources. Figure\ \ref{fig1} shows the typical $\gamma$-ray add-back spectrum from all the clover detectors measured at E$_{\textrm{beam}}$ = 38 MeV for $^{7}$Li+$^{205}$Tl system. The $\gamma$ lines from the possible evaporation residues ($^{208,209}$Po) following CF are labeled. Also the $\gamma$ lines from the residues ($^{207,208}$Bi) following the $\alpha$ capture channel and t capture channel ($^{206,207}$Pb) are marked.

\section{\label{sec:Result} Results and Discussion}
\subsection{Data reduction}
Data reduction procedure is similar as discussed in detail in our previous works \cite{vvp10, VVP18, VVP18b, Pandit17}. The cross sections for all the residues formed in CF ($^{208,209}$Po) and ICF ($^{207,208}$Bi, $^{206,207}$Pb) were determined considering all the ground and metastable ($\sim$ few $\mu$s life times) states. The $\gamma$ transitions populating the ground and metastable states in these nuclei are taken from Refs.\ \cite{Poletti1997,Poletti2000,Lonnroth1978,Fornal2003,Poletti1994,Hausser1972}. The cross sections for identified $\gamma$ transitions were calculated from the formula \[ 
\sigma _\gamma = \frac{{Y_\gamma  }}{{Y_M }} \frac{{d\Omega _M }}{{\epsilon_\gamma  }} \frac{{d\sigma_{Ruth}}}{d\Omega}, 
\] where $Y_\gamma$ is the $\gamma$-ray yield considering the internal conversion, $Y_M$ is the  elastic yield at the monitor detector, $d\Omega _M$ is the solid angle of the monitor detector, $\epsilon_\gamma$ is the absolute efficiency of the detection system for a specific $\gamma$ ray energy, and $\frac{{d\sigma_{Ruth}}}{d\Omega}$ is the Rutherford cross section (at $\theta_M$ = 25$^\circ$) at the same beam energy. \color{black} For the non even-even residues, the direct feeding to the ground state will not emit any $\gamma$-ray and hence not detected, however it is expected to be substantially small as observed in various studies \cite{Mukh06, Pradhan11, Brod75, vvp10}. \color{black} 

\begin{figure}
\includegraphics[width=0.54\textwidth,trim=1cm 17.2cm 6.0cm 3cm, clip=true]{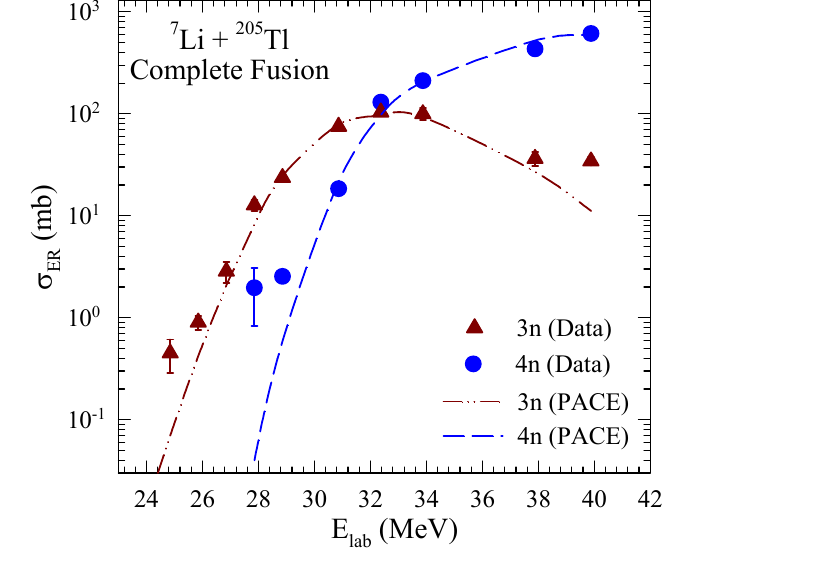}
\caption{\label{7Li_ER} ER cross sections for 3n ($^{209}$Po) and 4n ($^{208}$Po) channels following CF are shown as triangle and circle respectively. The results of the statistical model calculations for the corresponding ERs are shown by dashed dot dot (3n) and long dashed (4n) lines.}
\end{figure}

The cross sections for $^{209}$Po (3n) and $^{208}$Po (4n) ERs following CF along with the statistical model predictions using {\sc PACE} code \cite{Gav80} are shown in Fig.~\ref{7Li_ER}. In the {\sc PACE} calculations, the cross section for each partial wave ($\textit{l}$ distribution) obtained from the Coupled Channel calculation code {\sc CCFULL} \cite{Hag99} were fed as an input. The default optical potentials available in the code were used. The only free parameter remaining in the {\sc PACE} input was the level density parameter `a', which showed a negligible dependence on the values between a = A/9 and a = A/10. The CF cross sections were determined by dividing the cumulative measured ($\sigma_{3n+4n}^{expt}$) cross sections by the ratio R, which gives the missing ER contribution, if any. Here the ratio R is defined as \begin{math}{\rm{R = }}\sum\limits_{\rm{x}} {{\rm{\sigma }}_{_{{\rm{xn}}} }^{{\rm{PACE}}} } {\rm{/\sigma }}_{_{{\rm{fus}}} }^{{\rm{PACE}}}\end{math}, where x = 3, 4. The ratio (R) and the CF cross sections thus obtained are listed in Table\ \ref{tab:fus}. The error bars on the data are due to errors in the determination of the $\gamma$-ray yields, background subtraction and absolute efficiency of the detectors.

\begin{table}[htbp]
\caption{Measured cross sections for $\Sigma\sigma_{xn}$(x = 3, 4) evaporation residues and complete fusion along with the ratio R, obtained from PACE (defined in the text) for $^7$Li+$^{205}$Tl system for the measured energy range.}
\begin{tabular}
{ccccc}
\hline
E$_{lab}$ & E$_{c.m.}$ & $\sigma_{3n+4n}^{expt}$ & R(PACE) & $\sigma_{CF}^{expt}$ \\
(MeV) & (MeV) & (mb) & & (mb) \\
\hline
24.9 & 24.0 & 0.45 $\pm$ 0.16 & 0.90 & 0.50 $\pm$ 0.18 \\
25.9 & 25.0 & 0.90 $\pm$ 0.15 & 0.93 & 0.96 $\pm$ 0.16 \\
26.9 & 26.0 & 2.85 $\pm$ 0.65 & 0.96 & 2.95 $\pm$ 0.68 \\
27.9 & 26.9 & 14.7 $\pm$ 1.9 & 0.98 & 15.0 $\pm$ 2.0 \\
28.9 & 27.9 & 26.2  $\pm$ 1.6 & 0.98 & 26.6 $\pm$ 1.7 \\
30.9 & 29.9 & 93.4 $\pm$ 6.4 & 0.99 & 94.5 $\pm$ 6.5 \\
32.4 & 31.3 & 234 $\pm$ 12 & 0.99 & 236 $\pm$ 13 \\
33.9 & 32.8 & 311 $\pm$ 20 & 0.98 & 316 $\pm$ 20 \\
37.9 & 36.6 & 470 $\pm$ 43 & 0.97 & 485 $\pm$ 45 \\
39.9 & 38.6 & 647 $\pm$ 60 & 0.89 & 729 $\pm$ 67 \\
\hline
\end{tabular}
\label{tab:fus}
\end{table}

\subsection{\label{sec:Coupled} Coupled Channel Calculations}
\begin{figure}
\includegraphics[width=0.54\textwidth,trim=1.0cm 17.3cm 8.0cm 3cm, clip=true]{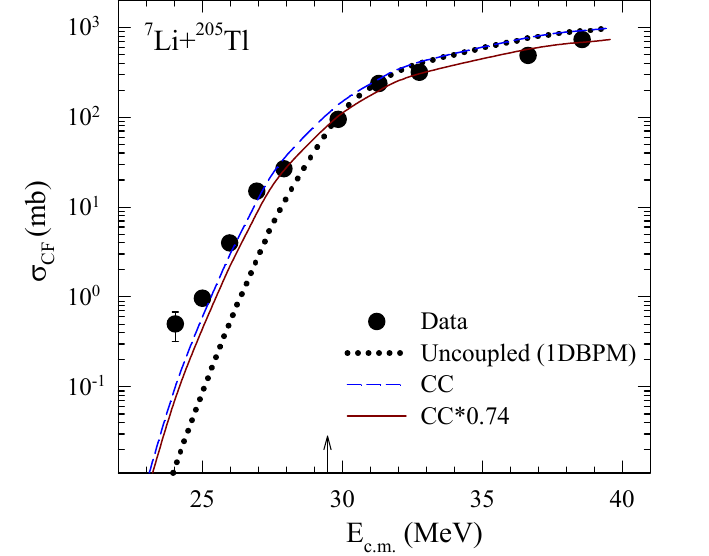}
\caption{\label{7Li_CF} Complete fusion cross section (filled circles) for the $^7$Li+$^{205}$Tl system compared with coupled (dashed lines) and uncoupled (dotted lines) results from CCFULL calculations. Solid lines were obtained by multiplying the coupled results by a factor of 0.74.}
\end{figure}

\begin{figure}
\includegraphics[width=0.54\textwidth,trim=0.2cm 19.0cm 9.5cm 1.5cm, clip=true]{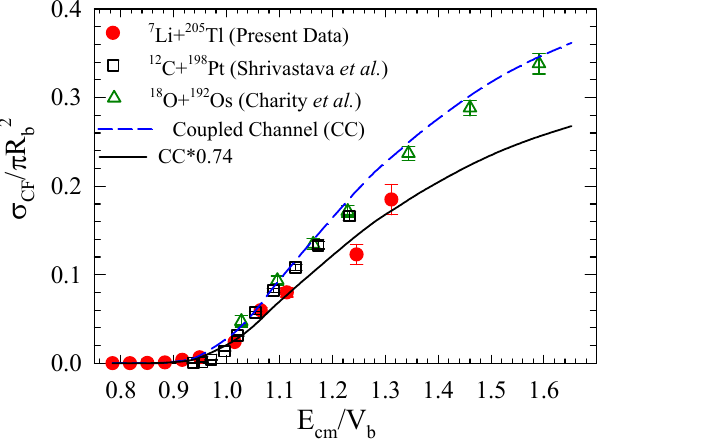}
\caption{\label{7Li_WBN_SBN_Comparision} Reduced cross sections as a function of reduced energy for the present system (filled circles) along with two other systems $^{12}$C+$^{198}$Pt (hollow square \cite{Shrivastava16}) and $^{18}$O+$^{192}$Os (hollow triangle \cite{Charity1986}). Dashed line is the result of coupled channel calculations \color{black} for $^7$Li+$^{205}$Tl system  \color{black}. Solid line is obtained by multiplying the coupled results by a factor of 0.74.}
\end{figure}
\begin{table}
\caption{Parameters for AW potential, along with the corresponding derived barrier height (V$_{b}$), barrier radius (R$_{b}$) and curvature ($\hbar\omega$)}
\begin{tabular}{ccccccc}
\hline \hline \\
System & V$_{0}$ & r$_{0}$ & a & V$_{b}$ & R$_{b}$ & $\hbar$$\omega$\\
&(MeV) & (fm) & (fm) & (MeV) & (fm) & (MeV) \\
\hline
$^{7}$Li+$^{205}$Tl & 47.4 & 1.18 & 0.63 & 29.4 & 11.2 & 4.46\\
$^{12}$C+$^{198}$Pt & 59.2 & 1.18 & 0.65 & 55.9 & 11.3 & 4.60\\
$^{18}$O+$^{192}$Os & 63.8 & 1.18 & 0.66 & 71.0 & 11.6 & 4.20\\
\hline
\end{tabular}
\label{tab:Pot}
\end{table}
Coupled channel calculations were performed using the modified version of CCFULL \cite{Hag99}, which can include the effect of projectile ground-state spin and its excitation in addition to the target excitation. The initial input potential parameters were obtained from the Woods-Saxon parameterization of the Aky$\ddot{\textrm{u}}$z-Winther (AW) potential \cite{Bro91} and are given in Table\ \ref{tab:Pot}. The table shows the corresponding uncoupled fusion barrier parameters (barrier height V$_b$, radius R$_b$, and curvature $\hbar\omega$). The full couplings include the coupling of the projectile ground state (3/2$^-$) and first excited state (1/2$^-$, 0.478 MeV) with $\beta_{00}$ ($\beta_2$ for the ground-state reorientation) = 1.189, $\beta_{01}$ ($\beta_2$ for the transition between the ground and the first excited states) = $\beta_{11}$ ($\beta_2$ for the reorientation of the 1st excited state) = 1.24. These values are taken from Ref.\ \cite{Beck03, Rath7Li}. These deformation parameters for $^7$Li are the same for both nuclear and Coulomb couplings. As the target is odd-A nucleus $^{205}$Tl, the excitation energies and deformation parameters were taken to be the averages of those of the neighbouring even-even nuclei $^{204}$Hg and $^{206}$Pb. The averaged 3$^{-}$ vibrational excited state with E$_x$ = 2.662 MeV, $\beta_3$ = 0.103 \cite{KIBEDI2002} was used. The effect of coupling of 2$^{+}$ excited state (E$_x$ = 0.620 MeV, $\beta_2$ = 0.05) \cite{RAMAN20011} is found to be less important compared to 3$^-$ state. The breakup or transfer coupling channel can not be included in these calculations.

The results from the uncoupled and coupled channel calculations are shown in Fig.~\ref{7Li_CF} by dotted and dashed lines, respectively. It was observed that at sub-barrier energies, the calculated fusion cross sections with the couplings (dashed lines) are enhanced compared to the uncoupled values. However, at above-barrier energies, the calculated values of fusion with or without couplings are higher than the measured ones. Agreement could be obtained when the calculated fusion cross sections are scaled by  a factor of 0.74, and the resulting scaled calculations are shown in Fig.~\ref{7Li_CF} by solid line. Thus, one can conclude that the CF cross sections in this region are suppressed by 26 $\pm$ 4\% compared to the prediction of CCFULL calculations. The uncertainty of 4\% in suppression factor was estimated from the uncertainties in V$_b$ and $\sigma_{CF}$. This suppression factor is similar to earlier studies with $^7$Li projectile \cite{Jha20,Kundu16,Wang14} on various targets which confirms that the CF suppression is target independent.

Present CF cross section data were also compared with the CF data for other systems forming similar compound nucleus (CN) but involving strongly bound projectiles. For the present data, the CN populated is $^{212}$Po, while with $^{12}$C+$^{198}$Pt \cite{Shrivastava16} and $^{18}$O+$^{192}$Os \cite{Charity1986}, the CN populated is $^{210}$Po. Figure \ref{7Li_WBN_SBN_Comparision} shows the comparison of the reduced cross sections as a function of reduced energy for the present system along with two other systems $^{12}$C+$^{198}$Pt \cite{Shrivastava16} and $^{18}$O+$^{192}$Os \cite{Charity1986}. It is interesting to see that the reduced fusion cross sections involving strongly bound projectiles are much larger than those for the weakly bound $^7$Li projectile, and they also agree with the results of coupled channel calculations without any scaling. However the coupled channel calculations multiplied by 0.74 match with the data for $^{7}$Li+$^{205}$Tl system. This again confirms the suppression of CF cross sections with weakly bound projectiles in comparison with strongly bound projectiles as well as those predicted by the fusion model adopted in CCFULL. Similar conclusion is also drawn in previous works \cite{Dasgupta04,Rath09,Rath12,Pals10}. 

\color{black}The coupled channels effects which lead to modification of barrier height are not so significant at above barrier energies. The calculated CF cross sections show a small modification due to the coupling effects and relatively independent of choice of standard potential parameters. A small suppression is found due to continuum-continuum couplings in coupled channels calculations \cite{Jha09}. In recent studies, it has been found that the clustering of the projectile and not the breakup in usual sense is responsible for the suppression \cite{Lei2019, Cook2018}. The clustering remains unaltered for a given projectile leading to the target independence in CF suppression. \color{black}

\subsection{\label{sec:ICF} ICF cross sections} \label{sec:ICF}
\begin{figure}[htbp]
\includegraphics[width=0.54\textwidth,trim=5.4cm 6.4cm 4.3cm 8.3cm, clip=true]{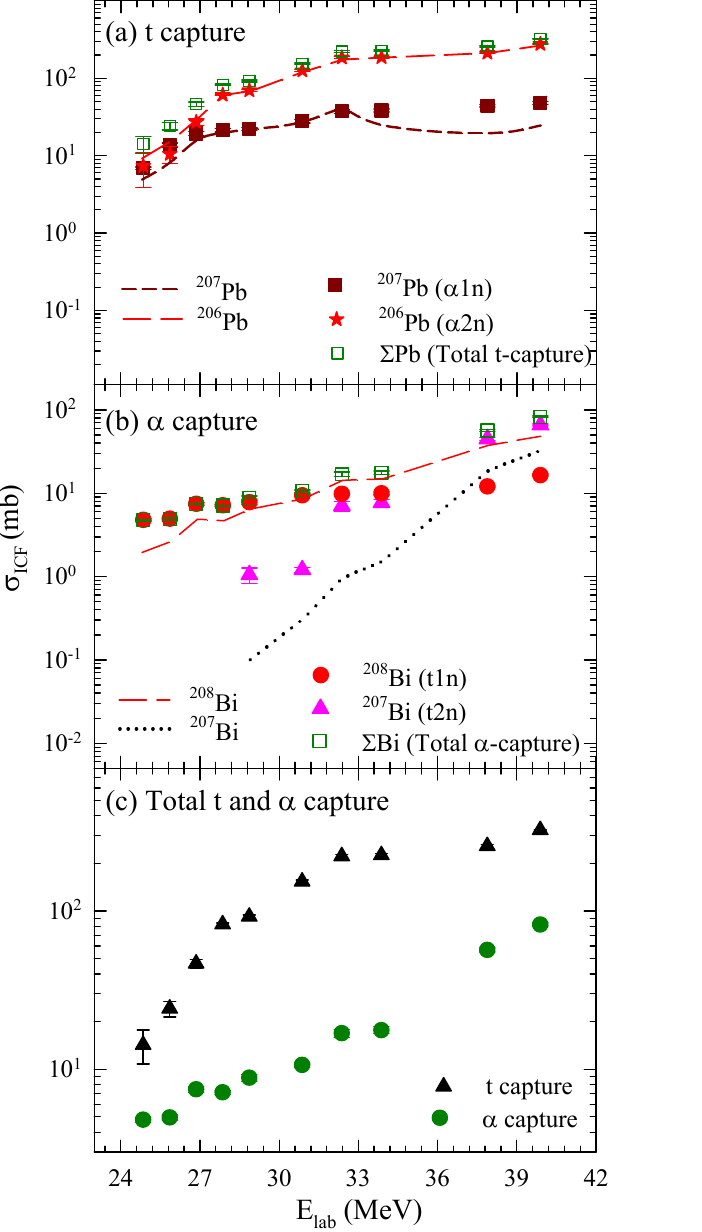}
\caption{\label{7Li_ICF}Measured residue cross sections for (a) t-capture and (b) $\alpha$-capture in $^{7}$Li+$^{205}$Tl system. The lines are the predictions from statistical model calculations for the corresponding residues. (c) total t-capture and $\alpha$-capture cross sections are compared. (see text for details).}
\end{figure}
\begin{table}
\caption{\label{tab:table4} Measured cross sections for ICF products obtained from online $\gamma$-ray measurement technique in $^7$Li+$^{205}$Tl system.}
\small
\begin{tabular}
{cccccccc}
\hline \hline \\
E$_{lab}$ & $^{208}$Bi & $^{207}$Bi & $^{207}$Pb & $^{206}$Pb  \\
(MeV) & (mb) & (mb) & (mb) & (mb) \\
\hline \\
24.9 & 4.8 $\pm$ 0.2 &  & 6.9 $\pm$ 0.3 & 7.4 $\pm$ 3.4  \\
25.9 & 4.9 $\pm$ 0.2 &  & 13.6$\pm$ 0.9 & 10.5 $\pm$ 2.6  \\
26.9 & 7.5 $\pm$ 0.3 &  & 19.4$\pm$ 1.2 & 27.1 $\pm$ 2.7  \\
27.9 & 7.5 $\pm$ 0.2 &  & 21.4$\pm$ 1.1 & 61.1 $\pm$ 1.5   \\
28.9 & 7.8 $\pm$ 0.4 & 1.1 $\pm$ 0.2 & 22.2$\pm$ 0.9 & 69.9 $\pm$ 2.2   \\
30.9 & 9.4 $\pm$ 0.3 & 1.2 $\pm$ 0.1 & 27.9$\pm$ 1.4 & 126 $\pm$ 4   \\
32.4 & 9.8 $\pm$ 0.3 & 7.1 $\pm$ 1.0 & 37.4$\pm$ 2.0 & 184 $\pm$ 5  \\
33.9 & 10.0 $\pm$ 0.5 & 7.7 $\pm$ 0.8 & 38.4$\pm$ 1.8 & 187 $\pm$ 5 \\
37.9 & 12.0 $\pm$ 0.4 & 44.7 $\pm$ 1.6 & 44.2$\pm$ 1.7 & 213 $\pm$ 5 \\ 
39.9 & 16.4 $\pm$ 0.5 & 65.6 $\pm$ 1.8 & 48.0$\pm$ 2.4 & 276 $\pm$ 4 \\
\hline \hline
\end{tabular}
\end{table}
The residues from incomplete fusion; $\textit{viz.}$, $^{207,208}$Bi (from $\alpha$-capture) and $^{206,207}$Pb (from t-capture) were also identified via $\gamma$ lines from Refs.\ \cite{Lonnroth1978, Fornal2003, Poletti1994, Hausser1972} and cross sections were extracted. The measured residue cross sections from t-capture and $\alpha$-capture are shown in Figs.\ \ref{7Li_ICF}(a) and \ref{7Li_ICF}(b), respectively and are listed in Table~\ref{tab:table4}. The total t-capture and total $\alpha$-capture cross sections are obtained from adding the individual residue cross sections and they are compared in Fig.~\ref{7Li_ICF}(c). The total t-capture cross sections are found to be much larger than $\alpha$-capture at all the measured energies. Similar observation is also reported in Refs.\ \cite{VVP18, VVP18b, Pandit17, VVP16}. It is to be noted that deuteron and proton stripping from $^7$Li projectile would give the same ERs as those following t-capture process and subsequent few neutron evaporation. Hence, from experiments it is difficult to separate these three processes.

In order to investigate the nature of t-capture and $\alpha$-capture cross sections in the $^7$Li+$^{205}$Tl system, statistical model calculations using PACE \cite{Gav80} code were performed. The code was modified accordingly for particle evaporations from the composite systems formed in fragment-capture reactions and used in previous works \cite{Sanatthesis, VVP18, VVP18b}. The spectrum of the surviving $\alpha$-particles, after capture of the complementary fragment (triton), represents the cross section for breakup-fusion as a function of the kinetic energy of the $\alpha$-particles. As seen from the literature \cite{Shrivastava13,Pfei73,Sign03,Santra12,Pradhan13,Jha20} for $^{6,7}$Li induced reactions on various targets, the $\alpha$, deuteron and triton energy spectra have typically 10 MeV  energy widths centered at $\frac{3}{7}$E$_{c.m.}$ and $\frac{4}{7}$E$_{c.m.}$ for t and $\alpha$ spectra, respectively. Two separate calculations assuming Gaussian distribution were performed. The calculated values of absolute cross sections for the residues from t-capture, $^{206,207}$Pb, are plotted as lines in Fig.~\ref{7Li_ICF}(a) showing reasonably good agreement with the data. Similarly, the calculations for $^{207,208}$Bi from $\alpha$-capture are plotted as lines in Fig.\ \ref{7Li_ICF}(b), showing a \color{black} reasonable \color{black} agreement. These calculations suggest that these residues are populated via fragment capture or cluster transfer followed by evaporation \cite{sanat21}.

\subsection{\label{sec:Coupled} Systematics of CF, ICF and TF cross sections with $^7$Li projectile}
\begin{figure}[htbp]
\includegraphics[width=76 mm,trim=6.0cm 7.5cm 6.5cm 0.4cm,clip=true]{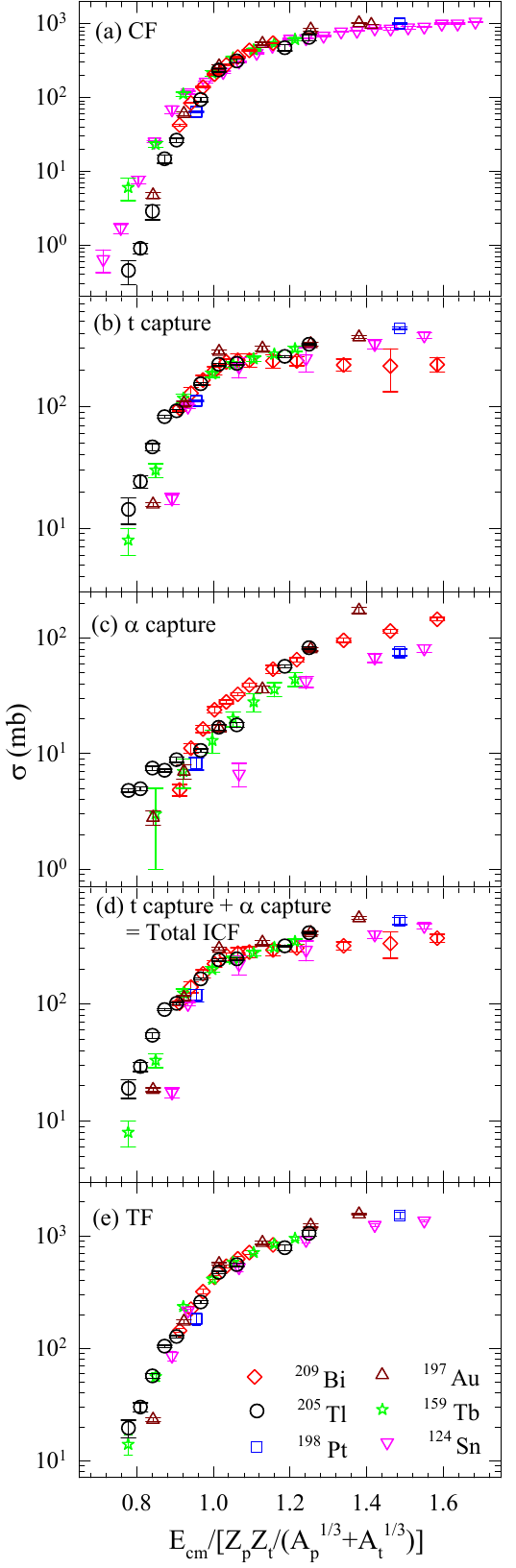}
\caption{\label{syst} Systematic behaviour of (a) CF, (b) t capture, (c) $\alpha$ capture, (d) Total ICF (= t capture + $\alpha$ capture) and (e) TF (= CF+ICF) cross sections as a function of reduced energy with $^7$Li+$^{124}$Sn \cite{VVP18}, $^{159}$Tb \cite{Brod75}, \color{black}$^{197}$Au \cite{Pals14}\color{black}, $^{198}$Pt \cite{Shrivastava13}, $^{209}$Bi \cite{Dasgupta04} and $^{205}$Tl (present data) systems.}
\end{figure}

\begin{figure} [htbp]
\includegraphics[width=70 mm,trim=1.7cm 0.2cm 11.4cm 0.6cm,clip=true]{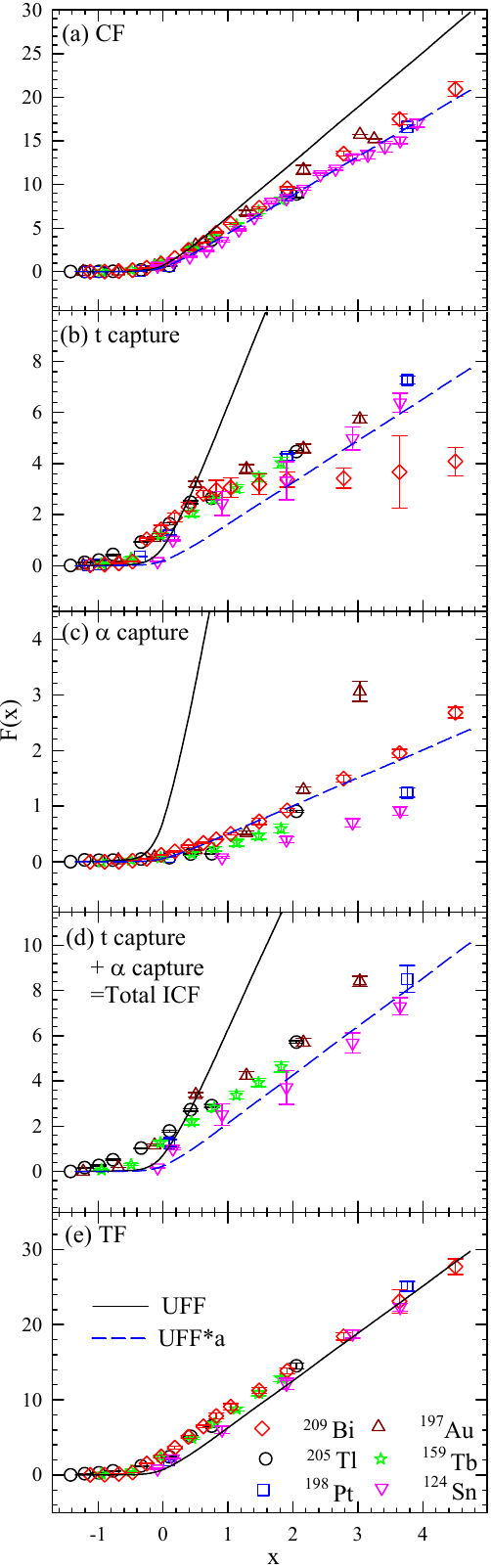}
\caption{\label{UFF} Measured fusion function F(x) for (a) CF, (b) t capture, (c) $\alpha$ capture, (d) Total ICF (= t capture + $\alpha$ capture) and (e) TF (= CF+ICF) for $^7$Li+$^{124}$Sn \cite{VVP18}, $^{159}$Tb \cite{Brod75}, \color{black}$^{197}$Au \cite{Pals14}\color{black}, $^{198}$Pt \cite{Shrivastava13}, $^{209}$Bi \cite{Dasgupta04} and $^{205}$Tl (present data) systems in linear scale. The solid and dashed line represents the UFF and UFF multiplied by a factor `a'. The value of 'a' is 0.70, 0.26, 0.08 and 0.34 for CF, t capture, $\alpha$ capture and total ICF respectively (see text for details).}
\end{figure}

The complete data set available for CF, $\alpha$ capture and t capture cross sections with $^7$Li projectile on $^{124}$Sn \cite{VVP18}, $^{159}$Tb \cite{Brod75}, \color{black}$^{197}$Au \cite{Pals14}\color{black}, $^{198}$Pt \cite{Shrivastava13}, $^{209}$Bi \cite{Dasgupta04} and $^{205}$Tl (present data) are utilised in the systematic plots shown in Figs.\ \ref{syst}(a-c) respectively. Total ICF is sum of $\alpha$-capture and t-capture cross sections and TF which is sum of CF and ICF cross sections are shown in Figs.\ \ref{syst}(d-e) respectively. The variable on x-axis (E$_{red}$) is chosen so as to remove any geometrical factors due to target size. This kind of comprehensive systematic including  for all channels simultaneously have been presented for the first time with the $^7$Li projectile. Similar plots were also shown earlier for the inclusive $\alpha$ \cite{Santra12, Pandit17, Jha20, VVP_EPJA23}, neutron transfer \cite{Kaushik21,Fang16,Parkar21,Parkar23}, fusion \cite{Jha20} and reaction \cite{Kola16} cross sections. As can be seen from the Figs.\ \ref{syst}(a-e), all the data are reasonably close in the range of 0.9-1.2 E$_{red}$, while it deviates below 0.9E$_{red}$ and above 1.2E$_{red}$. \color{black} It is to be noted that, the t capture data for $^7$Li+$^{209}$Bi system and $\alpha$ capture data for $^7$Li+$^{124}$Sn above 1.1 V$_b$ are deviating from the systemtic trend. \color{black}

Further to rule out the dependence on potential parameters, a reduction procedure was adopted \cite{Canto_UFF, Canto_UFF2} that completely eliminates the geometrical and static effects of the potential acting between the interacting partners. Thus, any deviation of the measured CF function from the universal fusion function (UFF) may be due to the breakup of the incident projectile. The fusion cross section and the incident energy are reduced to a dimensionless equation called the fusion function $F_{i}(x)$ and dimensionless variable \textit{x} in this reduction method:
\begin{equation*}
{F_{i}(x)= \frac{2E_{cm}}{\hbar\omega R_{b}^{2}}}\sigma_{i}~~~  \textrm{and} ~~~
x=\frac{E_{cm}-V_{b}}{\hbar\omega}\\
\end{equation*}
where, \textit{i} represents CF, t-capture, $\alpha$-capture, Total ICF and TF\color{black}\\

Potential parameters \textit{R$_{b}$}, \textit{V$_{b}$}, and $\hbar \omega$ were derived from AW potential parameters for various systems considered from Ref.\ \cite{NRV} to deduce \textit{F}(\textit{x}) and \textit{x}. The experimental CF, ICF and TF functions were reduced to respective fusion functions derived from the Wong formula \cite{Wong73}. After simplification of the Wong formula, F(x) reduces to
\begin{equation*}
F_{0}(x)= \textstyle\ln [1+exp(2\pi x)]\\ 
\end{equation*}
which is known as the universal fusion function (UFF). This function is valid for the above barrier data only. Respective fusion functions for CF, t-capture, $\alpha$-capture, Total ICF and TF are plotted in Fig.\ \ref{UFF}(a-e) in linear scale. In these plots, UFF is shown by solid line while UFF multiplied by a factor `a' is shown by dashed line. The `a' parameter is the fitting parameter to fit the corresponding data in (a-d). As can be seen, from Fig.\ \ref{UFF}(a), for CF data, the UFF multiplied by 0.7 (dashed blue line) match with the data. Thus, it can be concluded from the UFF calculations that the measured CF functions are $\sim$ 30\% less as compared to UFF, which is in agreement with the deduced suppression value of $\sim$ 26 $\pm$ 4\% within the error bars. Similarly, for t-capture  (Fig.\ \ref{UFF}(b)) and $\alpha$-capture  (Fig.\ \ref{UFF}(c)) data, the UFF multiplied by 0.26 and 0.08 respectively fairly match with the data. \color{black} The t-capture for $^7$Li+$^{209}$Bi system and $\alpha$ capture for $^7$Li+$^{124}$Sn above 1.1 V$_b$ are deviating much from the multiplied UFF line. This may be due to measurement error in the reported data \cite{Dasgupta04, VVP18}. \color{black} For the total ICF  (Fig.\ \ref{UFF}(d,i)), which is sum of t-capture and $\alpha$-capture cross sections, the UFF multiplied by 0.34 reasonably match with the data. This also shows that ICF contributions account for the suppression observed in the CF cross sections. Finally, the TF, which is the  sum of CF and ICF cross sections, reasonably agree with the UFF as shown in Fig.\ \ref{UFF}(e). 

\section{\label{sec:Sum} Summary}
Excitation functions for the complete and incomplete fusion of $^7$Li+$^{205}$Tl system were measured in the energy range 0.80 $<$ V$_b$ $<$ 1.34 by online $\gamma$-ray measurement technique. At above barrier energies, the measured CF cross sections were found to be suppressed by a factor of 26 $\pm$ 4\% in comparison with the coupled channel calculations, which is in agreement with the literature data for the $^7$Li projectile on various targets. A comparison of the CF cross sections for the present system with other systems involving strongly bound projectiles such as $^{12}$C+$^{198}$Pt and $^{18}$O+$^{192}$Os forming similar compound nuclei, clearly shows that the CF cross sections for the present system are systematically lower at above barrier energies. The measured t capture cross sections are significantly more than the $\alpha$ capture cross sections at all the energies. A systematic comparison of measured CF, ICF and TF data with $^7$Li projectile is shown. \color{black}The ground state Q values for triton transfer are positive while they are negative (except for $^7$Li+$^{124}$Sn system) for $\alpha$ transfer. The former is more dominant of the cluster transfers \cite{Jha20, VVP18, VVP18b}. However there is essentially no change in the Q$_{opt}$ values for these systems. It is around 0.68$\times$E$_{cm}$ for triton transfer while 0.34$\times$E$_{cm}$ for $\alpha$ transfer. This explains rather constant behaviour of cluster t-transfer  which is responsible for fusion suppression as argued above. \color{black}

\begin{acknowledgments}
The authors would like to thank the Pelletron-Linac accelerator staff for the smooth operation of the accelerator during the experiment. We also thank Mr. P. Patale for help during the experiment. P.M. acknowledges the financial support of Board of Research in Nuclear Science (BRNS), India (Sanction No: 58/14/04/2019-BRNS/10254) and CSIR-UGC, India in carrying out these investigations. P.M. and R.R. thank BARC for providing the facility to carryout the research work.

\end{acknowledgments}


\end{document}